\documentclass[12pt]{iopart_mod}
\usepackage{graphics}
\usepackage{bbm}
\usepackage{amssymb}
\usepackage{iopams}

\def\empile#1\over#2{\mathrel{\mathop{\kern 0pt#1}\limits_{#2}}}
\newcommand{\slvarepsilon}{\raise.15ex\hbox{$/$}\kern-.53em\hbox{$\varepsilon$}}
\newcommand{\slL}{\raise.15ex\hbox{$/$}\kern-.53em\hbox{$L$}}
\newcommand{\slP}{\raise.15ex\hbox{$/$}\kern-.53em\hbox{$P$}}
\newcommand{\slD}{\raise.15ex\hbox{$/$}\kern-.53em\hbox{$D$}}
\newcommand{\slp}{\raise.1ex\hbox{$/$}\kern-.63em\hbox{$p$}}
\newcommand{\slq}{\raise.1ex\hbox{$/$}\kern-.53em\hbox{$q$}}
\newcommand{\slv}{\raise.1ex\hbox{$/$}\kern-.63em\hbox{$v$}}
\newcommand{\slR}{\raise.15ex\hbox{$/$}\kern-.53em\hbox{$R$}}
\newcommand{\slQ}{\raise.15ex\hbox{$/$}\kern-.53em\hbox{$Q$}}
\newcommand{\slK}{\raise.15ex\hbox{$/$}\kern-.53em\hbox{$K$}}
\newcommand{\slk}{\raise.15ex\hbox{$/$}\kern-.53em\hbox{$k$}}
\newcommand{\slSigma}{\raise.15ex\hbox{$/$}\kern-.53em\hbox{$\Sigma$}}
\newcommand{\slcalP}{\raise.15ex\hbox{$/$}\kern-.63em\hbox{$\cal P$}}
\newcommand{\slA}{\raise.15ex\hbox{$/$}\kern-.73em\hbox{$A$}}
\newcommand{\slbfA}{\raise.15ex\hbox{$/$}\kern-.73em\hbox{${\imb A}$}}
\newcommand{\slpartial}{\raise.15ex\hbox{$/$}\kern-.53em\hbox{$\partial$}}
\newcommand{\sla}{\raise.15ex\hbox{$/$}\kern-.53em\hbox{$a$}}
\newcommand{\slb}{\raise.15ex\hbox{$/$}\kern-.53em\hbox{$b$}}
\newcommand{\slc}{\raise.15ex\hbox{$/$}\kern-.53em\hbox{$c$}}
\newcommand{\slC}{\raise.15ex\hbox{$/$}\kern-.63em\hbox{$C$}}

\def\P{{\boldsymbol P}}

\def\x{{\boldsymbol x}}
\def\y{{\boldsymbol y}}

\def\bs{\boldsymbol}

\catcode`\@=11

\newcount\@tempcntc
\def\@citex[#1]#2{\if@filesw\immediate\write\@auxout{\string\citation{#2}}\fi
  \@tempcnta\z@\@tempcntb\m@ne\def\@citea{}\@cite{%
        \@for\@citeb:=#2\do%
    {\@ifundefined{b@\@citeb}%
        {\@citeo\@tempcntb\m@ne\@citea%
                \def\@citea{,\penalty\@m\ }{\bf ?}\@warning%
                {Citation `\@citeb' on page \thepage \space undefined}}%
        {\setbox\z@\hbox{\global\@tempcntc0\csname b@\@citeb\endcsname\relax}
     \ifnum\@tempcntc=\z@ \@citeo\@tempcntb\m@ne%
       \@citea\def\@citea{,\penalty\@m}%
       \hbox{\csname b@\@citeb\endcsname}%
     \else%
      \advance\@tempcntb\@ne%
      \ifnum\@tempcntb=\@tempcntc%
      \else\advance\@tempcntb\m@ne\@citeo%
      \@tempcnta\@tempcntc\@tempcntb\@tempcntc\fi\fi}}\@citeo}{#1}}%

\def\@citeo{\ifnum\@tempcnta>\@tempcntb\else\@citea
  \def\@citea{,\penalty\@m}%
  \ifnum\@tempcnta=\@tempcntb\the\@tempcnta\else
   {\advance\@tempcnta\@ne\ifnum\@tempcnta=\@tempcntb \else
\def\@citea{--}\fi
    \advance\@tempcnta\m@ne\the\@tempcnta\@citea\the\@tempcntb}\fi\fi}

\catcode`\@=12

\begin{document}

\title[QCD at small $x$ and nucleus-nucleus collisions]{QCD at small $\x$ and nucleus-nucleus collisions}

\author{F Gelis}

\address{CEA/DSM/SPhT, Saclay\\
91191, Gif-sur-Yvette cedex, France}
\ead{francois.gelis@cea.fr}

\begin{abstract}
At large collision energy $\sqrt{s}$ and relatively low momentum
transfer $Q$, one expects a new regime of Quantum Chromo-Dynamics
(QCD) known as ``saturation''. This kinematical range is characterized
by a very large occupation number for gluons inside hadrons and
nuclei; this is the region where higher twist contributions are as
large as the leading twist contributions incorporated in collinear
factorization. In this talk, I discuss the onset of and dynamics in
the saturation regime, some of its experimental signatures, and its
implications for the early stages of Heavy Ion Collisions.
\end{abstract}

\section{Hadrons and nuclei at high energy}
A nucleon at low energy can be seen as made of three valence quarks,
constantly interacting via gluon exchanges and virtual fluctuations at
all space-time scales smaller than the size of the nucleon itself. In
an interaction process with a probe, only those fluctuations which are
longer lived/larger than the resolution of the probe are actually
relevant. In addition, interaction processes at low energy are made
very complicated by the fact that the constituents of the nucleon can
interact during the time seen by the probe. When one boosts the
nucleon to a higher energy, all its internal time scales are dilated,
which simplifies the interaction process with the probe\footnote{We
assume that we are in a frame in which the probe has not changed}: the
interactions among the constituents of the nucleon now occur over much
larger time-scales, and therefore the probe sees only a collection of
{\sl free} constituents. Moreover, the life-time of the quantum
fluctuations is also time dilated, and thus the number of gluons
taking part to the interaction process increases with the collision
energy. Simultaneously, the fluctuations that were already important
at the lower energy are now evolving so slowly that they can be
considered static over the time-scale seen by the external probe.

\begin{figure}[htbp]
\centerline{
\resizebox*{!}{4.5cm}{\includegraphics{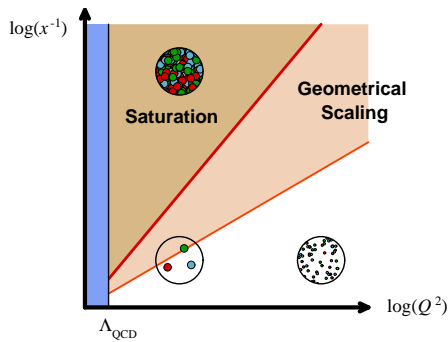}}
}
\caption{\label{fig:satdomain}Saturation domain in the $(Q^2,x)$
plane. We have also represented the larger domain where geometrical
scaling occurs.}
\end{figure}
However, this growth with energy of the number of gluons in the
wave-function of a nucleon (or nucleus) cannot continue
indefinitely. Indeed, it would imply that nucleon-nucleon
cross-sections grow faster than what is allowed by Froissart's
unitarity bound. In fact, an important aspect of the physics is
missing in the above picture: gluons can recombine when their
occupation number is large, a process known as {\sl gluon saturation}
\cite{saturation}. To quantify when this new phenomenon occurs, one
must compare the number of gluons per unit of transverse
area\footnote{This is the relevant density at high energy. Thanks to
Lorentz contraction in the longitudinal direction, soft gluons
belonging to different nucleons have overlapping wave-functions and
act coherently.}, $\rho\sim xG(x,Q^2)/\pi R^2$, and the cross-section
for recombination, $\sigma\sim\alpha_s/Q^2$. Saturation occurs when
$1\lesssim\rho\sigma$, or equivalently $Q^2\le Q_s^2(x)$, where
$Q_s(x)=\alpha_s xG(x,Q_s^2)/\pi R^2$ is known as the {\sl saturation
momentum}. The equation $Q^2=Q_s^2(x)$ delineates the border of the
saturation domain, indicated in figure
\ref{fig:satdomain}. Phenomenologically, $Q_s^2$ varies
like\footnote{The $A$ dependence follows from the fact that all the
gluons at a given impact parameter act coherently, and the $x$
dependence can be inferred from the gluon distribution measured at
HERA} $A^{1/3} x^{-0.3}$ with the nucleus atomic number $A$ and the
momentum fraction $x$.

\section{Color Glass Condensate}
The Color Glass Condensate (CGC) is a description of the nature of the
saturated nuclear matter, in which a saturation scale emerges
naturally. In the CGC description, the fast partons (large $x$) are
described as static color sources -- represented by a density
$\rho(\x_\perp)$ in the transverse plane -- that act as external
sources. Conversely, the slow partons (low $x$), dominated by gluons,
are described by the usual dynamical gauge fields $A^\mu$.  Such a
separation of the degrees of freedom was first proposed in the
McLerran-Venugopalan model \cite{MV}. In order to make this
description complete, one needs to specify the distribution
$W_{_Y}[\rho]$ for the hard sources in a projectile that has evolved
up to the rapidity $Y\equiv\ln(1/x)$. This distribution is in
principle non-perturbative, but its evolution with rapidity $Y$ is
controlled by an evolution equation that can be derived in
perturbation theory \cite{JIMWLK},
\begin{equation}
\frac{\partial W_{_Y}[\rho]}{\partial Y}
={\cal H}[\rho]\;W_{_Y}[\rho]\; ,
\end{equation}
known as the JIMWLK equation. This functional evolution equation can
be seen as an extension of the BFKL equation \cite{BFKL} that
incorporates the effect of gluon recombination. It has a very useful
mean field approximation: the Balitsky-Kovchegov equation
\cite{BK}. These evolution equations are presently known at leading
logarithmic accuracy, but several recent works have extended them in
order to include the Next-to-Leading Order corrections that come from
the running of the strong coupling constant
\cite{EvolRunning}. Another direction actively pursued is aimed at
including ``pomeron loops'' in this description (see
\cite{PomeronLoops2006} for recent advances). Pomeron loops can be
seen as fluctuations that are important in the region when the gluon
occupation number is low, and may affect crucially the rapidity
evolution of scattering amplitudes since it is controlled by the tail
of the gluon distribution.

\section{Experimental evidence of saturation}
\begin{figure}[htbp]
\centerline{
\hfil
\resizebox*{!}{4cm}{\includegraphics{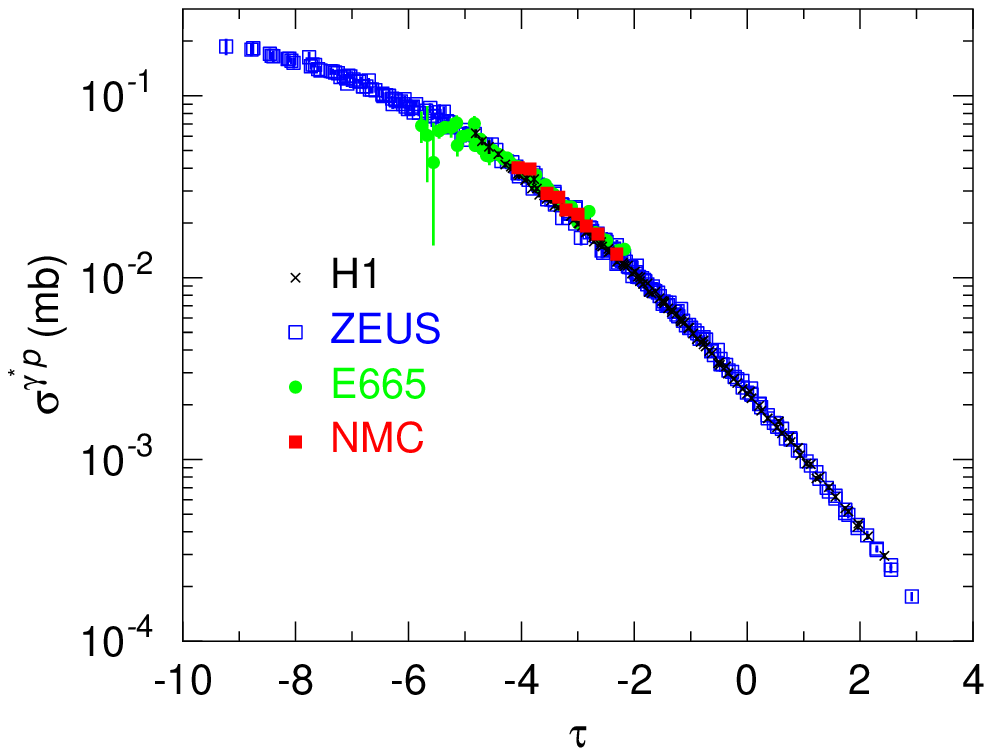}}
\hfil\hglue 10mm
\resizebox*{!}{4cm}{\includegraphics{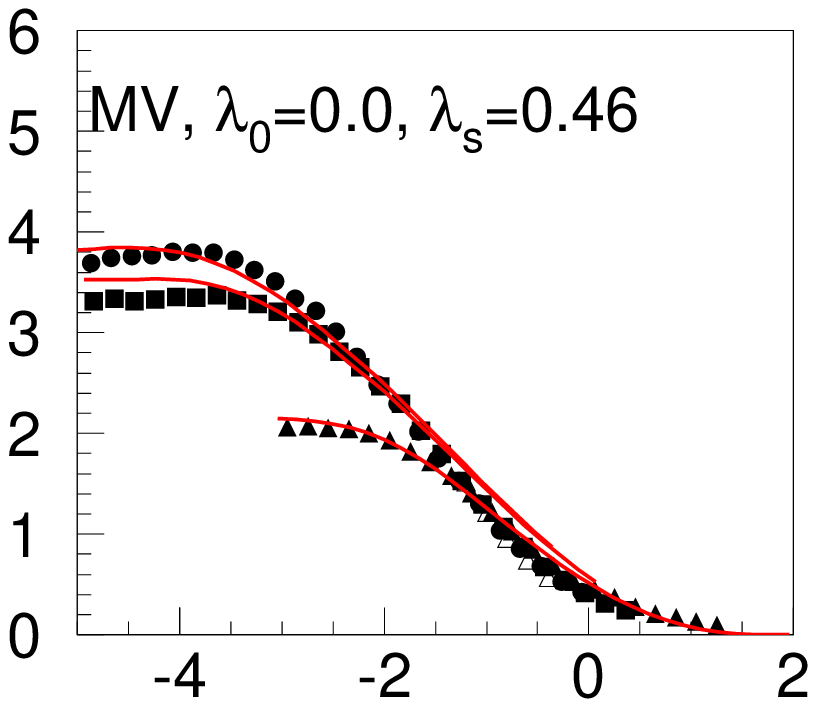}}
\hfil
}
\caption{\label{fig:GS-LF} Left: geometrical scaling in the
$\gamma^*$-proton cross-section at HERA. Right: Limiting fragmentation
in RHIC data.}
\end{figure}
The first evidence of saturation comes from the ``geometrical
scaling'' observed in Deep Inelastic Scattering at HERA
\cite{GS}. There, the measured $\gamma^*p$ cross-section at $x\le
0.01$ -- which in principle depends on both variables $x$ and $Q^2$ --
was found to depend only on the combination $\tau\equiv
\ln(Q^2/Q_s^2(x))$ with $Q_s^2\propto x^{-0.3}$ (see the left panel of
figure \ref{fig:GS-LF}). Such a scaling is a direct consequence of the
behavior of solutions of the evolution equation for the distribution
of sources in the target proton, and the value of the exponent $0.3$
can be accounted for from a careful analysis of these solutions
\cite{GS1}. Going beyond this simple scaling, saturation physics leads
to very good fits to a broad set of data from HERA and
RHIC \cite{fits}.

Another experimental result which is suggestive of saturation is the
so-called ``limiting fragmentation'' (right panel of figure
\ref{fig:GS-LF}). By shifting the rapidity axis by the beam rapidity,
one can see that the rapidity distribution for produced particles at
collisions of various energies tend to some universal curve in the
fragmentation region \cite{LF}. In the CGC framework, this property
follows naturally from the unitarization of scattering amplitudes in
the dense target, and the approximate Bjorken scaling in the
fragmenting nucleus. The limiting curve then appears to be a
reflection of the parton distribution at large $x$ in the fragmenting
nucleus \cite{LF1}.

\begin{figure}[htbp]
\centerline{
\resizebox*{!}{3.5cm}{\includegraphics{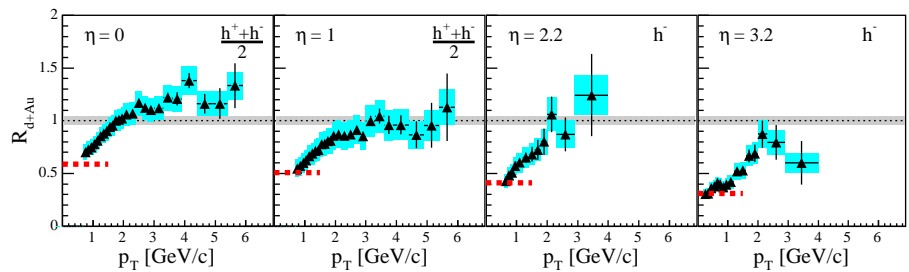}}
}
\caption{\label{fig:forward}Forward suppression observed by BRAHMS.}
\end{figure}
The observation of the suppression of hadron spectra at rather large
$p_\perp$ and forward rapidity in deuteron-gold collisions at RHIC
\cite{Forward} (see figure \ref{fig:forward} for BRAHMS results)
provides further information on the onset of saturation. The fact that
the nuclear modification factor $R_{\rm dA}$ is larger than unity at
mid rapidity is believed to be an effect of multiple scatterings
(Cronin effect), and the suppression of this ratio at forward
rapidities is a consequence of the shadowing that builds up via the
evolution in $Y$ \cite{Forward1}.

\section{Initial particle production in nucleus-nucleus collisions}
\begin{figure}[htbp]
\centerline{
\hfil
\resizebox*{!}{3.3cm}{\includegraphics{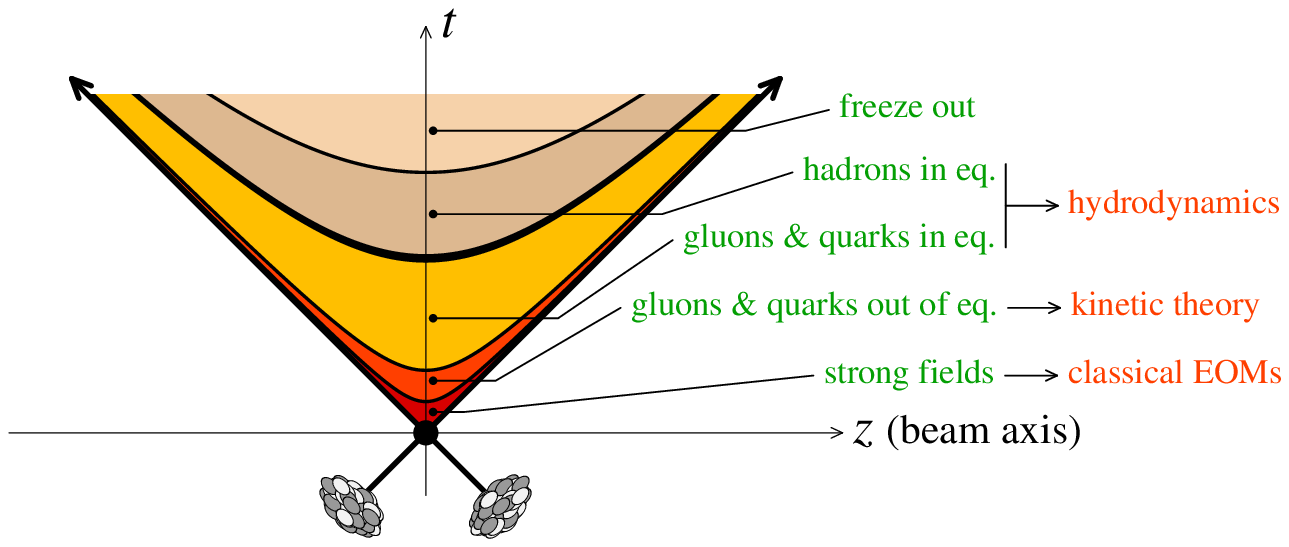}}
\hfil\hglue 5mm
\resizebox*{!}{3.3cm}{\includegraphics{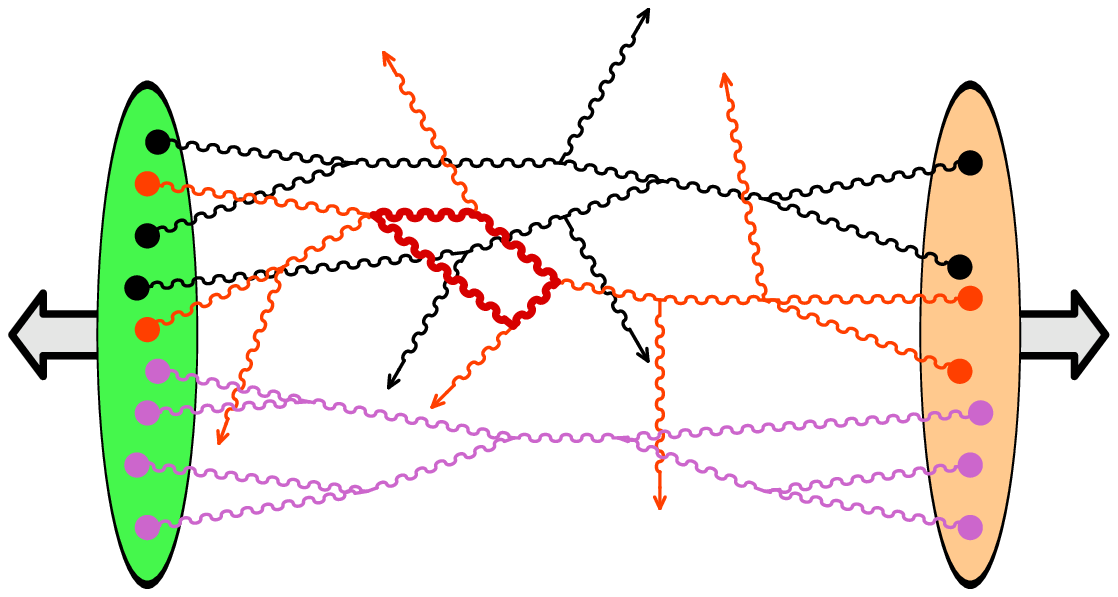}}
\hfil}
\caption{\label{fig:coll}Left: the successive stages of a heavy ion
collision. Right: typical diagram contributing to gluon production in
the collision of two nuclei at high energy.}
\end{figure}

In order to use the CGC framework to describe the collision of two
nucleons or nuclei, one needs two color sources $\rho_{1,2}$,
representing respectively the two projectiles, that couple to the
gauge fields (see \cite{H} for a detailed discussion on this) via the
following current,
\begin{equation}
J^\mu\equiv \delta^{\mu+}\delta(x^-)\rho_1(\x_\perp)+\delta^{\mu-}\delta(x^+)\rho_2(\x_\perp)\; .
\end{equation}
As illustrated in figure \ref{fig:coll}, the description of Heavy Ion
Collisions is usually split into several stages, each of them being
described by different theoretical tools. It is expected that
perturbation theory is appropriate for the very early stages, during
which the constituents of the incoming nuclei are released and
particles are initially produced.
However, in a typical central heavy ion collision at high energy, 99\%
of the produced particles have a transverse momentum below 2~GeV --
i.e. below the saturation momentum. In this region, collinear
factorization breaks down, and it is believed that the CGC framework
is more appropriate for describing the initial particle production.

In the saturation regime, the sources $\rho_{1,2}$ are typically of
order of $1/g$, which alters the power counting for gluon production:
adding an extra source to a given diagram does not change its order,
implying that the color sources $\rho_{1,2}$ should be included to all
orders. A typical diagram contributing to gluon production from these
sources is illustrated in the right panel of figure \ref{fig:coll}. In
general, there are several disconnected subdiagrams, including some
that do not produce any gluon (vacuum diagrams) \cite{GV}.

\subsection{Gluon production at Leading Order}
\begin{figure}[htbp]
\centerline{
\resizebox*{!}{4cm}{\includegraphics{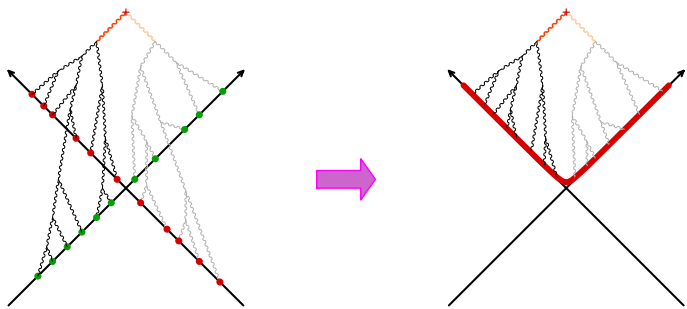}}
}
\caption{\label{fig:gluonLO}Gluon multiplicity at leading order. Left:
tree retarded diagrams contributing to the gluon spectrum. The color
sources are represented by red and green dots. Right: reformulation as
an initial value problem on the light-cone.}
\end{figure}
At leading order \cite{GluonLO}, one must sum all the tree diagrams,
with an arbitrary number of color sources -- in general a very
complicated task. However, for the inclusive gluon spectrum, this can
be done from the classical solution of Yang-Mills equations with {\sl
retarded boundary conditions}\footnote{Less inclusive observables can
be expressed at LO in terms of classical solutions with more
complicated boundary conditions \cite{GV}.}. The retarded nature of
the boundary conditions leads to an important simplification: one can
reformulate the problem as an initial value problem where one
specifies the fields and their first time derivatives just above the
light-cone (and these initial conditions can be obtained
analytically). This classical solution of the equation of motion is a
sum of tree diagrams with retarded propagators, as illustrated in
figure \ref{fig:gluonLO}. An important feature of the classical
solution of Yang-Mills equations is its invariance under boosts in the
longitudinal solution, which implies that the field has only zero
modes with respect to the rapidity variable $\eta$.

\subsection{Gluon production at Next to Leading Order}
In order to make the foundations of this description more robust, it
is important to study loop corrections\footnote{Quark production --
also a 1-loop contribution -- has been evaluated in
\cite{Quarks}.}. One reason for doing so is to check the factorization
of the leading logarithms of $1/x$ in the evolution of the
distribution of sources $W[\rho_{1,2}]$, which is necessary for the
internal consistency of this framework. Another motivation comes from
the recent observation that the LO boost invariant solution is
unstable against rapidity dependent perturbations \cite{RV} -- and
loop corrections are a natural origin for such perturbations.

\begin{figure}[htbp]
\centerline{
\resizebox*{!}{4cm}{\includegraphics{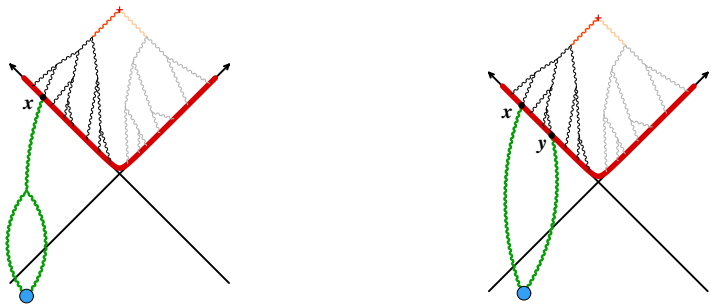}}
}
\caption{\label{fig:gluonNLO} The two contributions to the gluon
multiplicity at NLO.}
\end{figure}

Although the LO calculation of the gluon spectrum is non-perturbative
since it requires to solve the Yang-Mills equation to all orders in
the sources $\rho_{1,2}$, it turns out that one can express the 1-loop
contributions as a perturbation of the initial value problem
encountered at LO. There are two contributions, illustrated in figure
\ref{fig:gluonNLO}.  The NLO correction can be obtained by the action
of some analytically calculable operator on the LO contribution, seen
as a functional of the fields and canonical momenta on the light-cone
\cite{GLV},
\begin{equation}
\delta N_{_{NLO}}=\Big[
\!\!\!\!
\int\limits_{\x\in{\rm light-cone}}
\!\!\!\!\!\!\!\!
\delta{\cal A}(\x){\bs T}_\x
+
\frac{1}{2}
\!\!\!\!
\int\limits_{\x,\y\in{\rm light-cone}}
\!\!\!\!\!\!\!\!
{\bs\Sigma}(\x,\y){\bs T}_\x {\bs T}_\y\Big]\; 
N_{_{LO}}[{\cal A}_{\rm in}[\rho_1,\rho_2]]\; .\;
\label{eq:NLO}
\end{equation}
In this formula, $\delta{\cal A}(\x)$ and ${\bs\Sigma}(\x,\y)$ are
respectively the 1-point and 2-point functions represented in green in
figure \ref{fig:gluonNLO} -- they are in principle {\sl calculable
analytically}. ${\bs T}_\x$ is the generator of shifts of the initial
condition at point $\x$ on the light-cone for the solution of
Yang-Mills equations\footnote{Eq.~(\ref{eq:NLO}) is only a sketch of
the actual formula. Indeed, $N_{_{LO}}$ depends on both the initial
fields ${\cal A}_{\rm in}$ and momenta ${\Pi}_{\rm in}$ on the
light-cone, and there should therefore be shift operators for both of
them.}, i.e. a functional derivative $\delta/\delta {\cal A}_{\rm
in}(\x)$.  
Eq. (\ref{eq:NLO}) is sufficient to discuss the structure
of singularities that arise at 1-loop~:
\begin{itemize}
\item[{\bf i.}~] the coefficients $\delta{\cal A}$ and ${\bs\Sigma}$
are infinite, due to divergences in integrations over rapidity
\cite{LM}.
\item[{\bf ii.}~] the quantity ${\bs T}_\x\,{\cal A}(\tau,\y)$
diverges when the time $\tau$ goes to infinity. This is due to the
instability of the boost invariant classical solution found in
\cite{RV}.
\end{itemize}

The first type of divergences is related to the evolution with
rapidity of the distributions $W[\rho_{1,2}]$. For the CGC framework
to be consistent, one should be able to absorb these divergences in
the rapidity dependence of $W[\rho_{1,2}]$ -- a statement very similar
in spirit to the factorization of collinear divergences in
conventional perturbative QCD. Another way to state the same problem
is to set fictitious boundaries in rapidity, $Y_0$ and $Y_0^\prime$,
between the observable (here the number operator) and the color
sources. Now, all integrations over rapidity in the observable are
over the finite range $[Y_0^\prime,Y_0]$, and thus lead to a finite
result. Thus, $\delta{\cal A}$ and ${\bs \Sigma}$ become finite, but
$Y_0$ and $Y_0^\prime$ dependent. Naturally, the distributions of
color sources, $W[\rho_{1,2}]$, must now be evolved from the beam
rapidities to $Y_0$ and $Y_0^\prime$ respectively. The
factorization alluded to before is equivalent to the requirement that
the $Y_0$ and $Y_0^\prime$ dependence cancel between the 1-loop
correction to the observable and the rapidity evolution of the
distributions of sources. For this cancellation to happen, one will
need to establish the following relation for these divergent terms
\cite{GLV}~:
\begin{equation}
\Big[\delta N_{_{NLO}}\Big]_{{\rm divergent}\atop{\rm coefficients}}
=
\Big[
(Y_0-Y){\cal H}^\dagger[\rho_1]+(Y-Y_0^\prime){\cal H}^\dagger[\rho_2]
\Big]\;
N_{_{LO}}[{\cal A}_{\rm in}]\; ,\;
\end{equation}
where ${\cal H}[\rho]$ is the same Hamiltonian as in the JIMWLK
evolution equation.

The second type of problem is due to an instability of the {\sl boost
invariant} solution of the classical Yang-Mills equations
\cite{RV}. 
\setcounter{footnote}{1} After resummation of these divergences to all
orders, the operator acting on $N_{_{LO}}$ in Eq.~(\ref{eq:NLO}) will
be replaced by some functional $Z[{\bs T}_\x]$. Formally, one can
perform a ``Fourier transform'' of this functional by introducing some
auxiliary field $a(\x)$ on the light-cone\footnote{Again, the symbol
$a(\x)$ is a shortcut for both the field and canonical momentum on the
light-cone.}, and then rewrite the contribution of the resummed
unstable modes as\footnote{The linear term $\delta{\cal A}(\x){\bs
T}_\x$ cannot trigger the instability because $\delta{\cal A}(\x)$ is
boost invariant. Thus, $Z[{\bs T}_\x]=Z[-{\bs T}_\x]$ and its Fourier
transform $\widetilde{Z}[a]$ -- the Wigner distribution of the initial
fluctuations -- is real.}
\begin{eqnarray}
\Big[\delta N_{_{NLO}}\Big]_{{\rm unstable}\atop{\rm modes}}
&=&
\int \big[D a(\x)\big]\;\widetilde{Z}[a]\;
\exp\Big\{i\int_\x a(\x){\bs T}_\x\Big\}\;
N_{_{LO}}[{\cal A}_{\rm in}]
\nonumber\\
&=&\int \big[D a(\x)\big]\;\widetilde{Z}[a]\;
N_{_{LO}}[{\cal A}_{\rm in}+a]\; .
\end{eqnarray} 
In other words, resumming the unstable modes can be done by solving
the classical Yang-Mills equations with a fluctuation $a(\x)$
superimposed to the boost invariant initial condition ${\cal A}_{\rm
in}$, with a distribution $\widetilde{Z}[a]$. This is the formal
justification for the fluctuations added to the initial conditions in
\cite{RV}. By a completely different approach \cite{FGM}, it appears
that the functional $\widetilde{Z}[a]$ is a Gaussian, suggesting that
the resummation of the unstable modes is simply an exponentiation.

\section{Conclusions}
QCD at small $x$, and in particular gluon saturation, has become an
important aspect of hadronic collisions at RHIC and LHC energies. In
nucleus-nucleus collisions, the CGC is the appropriate framework to
study the early stages of the collision, because most of the particles
are produced from low $x$ gluons. It seems now technically feasible to
compute the 1-loop corrections to the gluon spectrum, and to resum
their diverging terms. Doing so would allow one to prove for the CGC
framework factorization results that are crucial for the overall
consistency of this approach.

\vglue 3mm
\noindent{\bf Acknowledgements~:}~ Important parts of this talk are
based on work done in collaboration with K. Fukushima, T. Lappi, L.
McLerran and R. Venugopalan.


\section*{References}
\begin{thebibliography}{10}

\bibitem{saturation}
{Gribov L V, Levin E M, Ryskin M G}, Phys. Rep. {\bf 100}, (1983) 1;
{Mueller A H, Qiu J-W}, \NP {\bf B268}, (1986) 427;
{Blaizot J P, Mueller A H}, \NP {\bf B289}, (1987) 847.


\bibitem{MV}
{McLerran L D, Venugopalan R}, \PR {\bf D49}, (1994) 2233; {\it ibid.} 3352;
{\it ibid.} {\bf D50}, (1994) 2225.


\bibitem{JIMWLK}
{Jalilian-Marian J, Kovner A, McLerran L D, Weigert H}, Phys. Rev. {\bf D55}, (1997) 5414;
{Jalilian-Marian J, Kovner A, Leonidov A, Weigert H}, Nucl. Phys. {\bf B504}, (1997) 415; Phys. Rev. {\bf D59}, (1999) 014014; {\it ibid.} 034007; {\it ibid.} Erratum 099903;
{Iancu E, Leonidov A, McLerran L D}, Nucl. Phys. {\bf A692},
  (2001) 583; Phys. Lett. {\bf B510}, (2001) 133;
{Ferreiro E, Iancu E, Leonidov A, McLerran L D}, Nucl. Phys. {\bf A703}, (2002) 489.

\bibitem{BK}
{Balitsky I}, Nucl. Phys. {\bf B463}, (1996) 99;
{Kovchegov Yu}, Phys. Rev. {\bf D61}, (2000) 074018.

\bibitem{BFKL}
{Balitsky I, Lipatov L N}, Sov. J. Nucl. Phys. {\bf 28}, (1978) 822;
{Kuraev E A, Lipatov L N, Fadin V S}, Sov. Phys. JETP {\bf 45}, (1977) 199.

\bibitem{EvolRunning}
Gardi E, Kuokkanen J, Rummukainen K, Weigert H, hep-ph/0609087;
Kovchegov Yu, Weigert H, hep-ph/0609090; hep-ph/0612071; 
Balitsky I, Phys. Rev. {\bf D75}, (2007) 014001.

\bibitem{PomeronLoops2006}
Hatta Y, Iancu E, Marquet C, Soyez G, Triantafyllopoulos D N, Nucl. Phys. {\bf A773}, (2006) 95;
Iancu E, Marquet C, Soyez G, Nucl. Phys. {\bf A780}, (2006) 52;
Blaizot J-P, Iancu E, Triantafyllopoulos D N, hep-ph/0606253;
Iancu E, de Santana Amaral J T, Soyez G, Triantafyllopoulos D N, hep-ph/0611105;
Shoshi  A I, Xiao B-W, hep-ph/0605282;
Bondarenko S, Motyka L, Mueller A H, Shoshi A I, Xiao B-W, hep-ph/0609213;
Kozlov M, Shoshi A I, Xiao B-W, hep-ph/0612053;
Kozlov M, Levin E, Nucl. Phys. {\bf A779}, (2006) 142;
Kozlov M, Levin E, Prygarin A, hep-ph/0606260;
Levin E, hep-ph/0608043;
Kozlov M, Levin E, Khachatryan V, Miller J, hep-ph/0610084;
Kovner A, Lublinsky M, Nucl. Phys. {\bf A779}, (2006) 220.


\bibitem{GS}
H1 collaboration, Acta Phys. Polon. {\bf B33}, (2002) 2841;
ZEUS collaboration, Phys. Lett. {\bf B345}, (1995) 576;
Stasto A M, Golec-Biernat K, Kwiecinski J, Phys. Rev. Lett. {\bf 86}, (2001) 596;
Gelis F, Peschanski R, Schoeffel L, Soyez G, hep-ph/0610435.

\bibitem{GS1}
Iancu E, Itakura K, McLerran L, Nucl. Phys. {\bf A708}, (2002) 327;
Munier S, Peschanski R, Phys. Rev. Lett. {\bf 91}, (2003) 232001; 
Triantafyllopoulos D N, Nucl. Phys. {\bf B648} (2003) 293. 

\bibitem{fits}
Iancu E, Itakura K, Munier S, Phys. Lett. {\bf B590}, (2004) 199;
Kowalski H, Motyka L, Watt G, Phys. Rev. {\bf D74}, (2006) 074016;
Forshaw J R, Sandapen R, Shaw G, JHEP {\bf 0611}, (2006) 025;
Dumitru A, Hayashigaki A, Jalilian-Marian J, Nucl. Phys. {\bf A770}, (2006) 57;
Goncalves V P, Kugeratski M S, Machado M V T, Navarra F S, Phys. Lett. {\bf B643}, (2006) 273. 

\bibitem{LF}
Bearden I G {\it et al}, Phys. Lett. {\bf B523}, (2001) 227; Phys. Rev. Lett. {\bf 88}, (2002) 20230;
Back B B {\it et al}, Phys. Rev. Lett. {\bf 91}, (2003) 052303; nucl-ex/0509034; 
Adams J {\it et al}, Phys. Rev. Lett. { \bf 95}, (2005) 062301; Phys. Rev. {\bf C73}, (2006) 034906.

\bibitem{LF1}
Jalilian-Marian J, Phys. Rev. {\bf C70}, (2004) 027902;
Bialas A, Je\.zabek M, Phys. Lett. {\bf B590}, (2004) 233;
Gelis F, Stasto A M, Venugopalan R, Eur. Phys. J {\bf C48}, (2006) 489.

\bibitem{Forward}
Arsene I {\it et al.}, Phys. Rev. Lett. {\bf 93} (2004) 242303;
Adams J {\it et al.}, Phys. Rev. Lett. {\bf 97}, (2006) 152302.

\bibitem{Forward1}
{Kharzeev D, Levin E, McLerran L D}, Phys. Lett. {\bf B561}, (2003) 93;
Baier R , Kovner A , Wiedemann U A, Phys. Rev. {\bf D68} (2003), 054009; 
Gelis F, Jalilian-Marian J, Phys. Rev. {\bf D67}, (2003) 074019;
Kharzeev D, Kovchegov Yu, Tuchin K, Phys. Rev. {\bf D68}, (2003) 094013; 
Phys. Lett. {\bf B599}, (2004) 23; 
{Albacete J L, Armesto N, Kovner A, Salgado C A, Wiedemann U A}, Phys. Rev. Lett. {\bf 92}, (2004) 082001;
{Iancu E, Itakura K, Triantafyllopoulos D N}, Nucl. Phys. {\bf A742}, (2004) 182;
{Blaizot J-P, Gelis F, Venugopalan R}, Nucl. Phys. {\bf A743}, (2004) 13;
Baier R, Mehtar-Tani Y, Schiff D, Nucl. Phys. {\bf A764}, (2006) 515.

\bibitem{H}
Hatta Y, Nucl. Phys. {\bf A781}, (2007) 104. 

\bibitem{GluonLO}
{Krasnitz A, Venugopalan R}, Nucl. Phys. {\bf B557}, (1999) 237;
                             Phys. Rev. Lett. {\bf 84}, (2000) 4309;
                             {\it ibid.} {\bf 86}, (2001) 1717;
{Krasnitz A, Nara Y, Venugopalan R}, Nucl. Phys. {\bf A727}, (2003) 427;
                                     Phys. Rev. Lett. {\bf 87}, (2001) 192302;
{Lappi T}, Phys. Rev. {\bf C67}, (2003) 054903.

\bibitem{GV}
Gelis F, Venugopalan R, Nucl. Phys. {\bf A776}, (2006) 135, {\it ibid.} {\bf A779}, (2006) 177.

\bibitem{Quarks}
Gelis F, Kajantie K, Lappi T, Phys. Rev. {\bf C71}, (2005) 024904; Phys. Rev. Lett. {\bf 96}, (2006) 032304.

\bibitem{RV}
Romatschke P, Venugopalan R, Phys. Rev. Lett. {\bf 96}, (2006) 062302; 
                             Eur. Phys. J. {\bf A29}, (2006) 71;
                             Phys. Rev. {\bf D74}, (2006) 045011;
 Venugopalan R, {\it these proceedings.}

\bibitem{GLV}
Gelis F, Lappi T, Venugopalan R, {\it work in progress.}

\bibitem{LM}
Lappi L, McLerran L D, Nucl. Phys. {\bf A772}, (2006) 200.


\bibitem{FGM}
Fukushima K, Gelis F, McLerran L D, hep-ph/0610416.

\end{thebibliography}
\end{document}